\documentclass[12pt,preprint]{aastex}
\usepackage{graphicx,epsfig,epsf,amssymb}
\def \<{\langle}
\def \>{\rangle}

\newcommand{\degree}{^\circ}

\begin{document}

\title{Inconsistency between WMAP Data and Released Map}

\author{Hao Liu\altaffilmark{1} and Ti-Pei Li\altaffilmark{1,2,3}}
 \altaffiltext{1}{Key Lab. of Particle Astrophys., Inst. of High Energy Phys.,
Chinese Academy of Sciences, Beijing}
 \altaffiltext{2}{Department of Physics \& Center for Astrophysics,
Tsinghua University, Beijing, China}
\altaffiltext{3}{Department of Engineering Physics \& Center for Astrophysics,
Tsinghua University, Beijing, China}

\begin{abstract}
A remarkable inconsistency between the calibrated 
differential time-ordered data (TOD) of the Wilkinson Microwave 
Anisotropy Probe (WMAP) mission,
which is the input for map-making, and the cosmic microwave background (CMB) 
temperature maps published by the WMAP team is revealed, indicating that 
there must exist a serious problem in the map making routine of the WMAP team. 
This inconsistency is easy to be confirmed without the use of WMAP map-making
software. In view of the importance of this issue for cosmology study, 
the authors invite readers to check it by themselves.
\end{abstract}

\section{INTRODUCTION}
The CMB data from the WMAP mission are the most important bases of
cosmology study, and the accuracy of the CMB map recovered from
the WMAP data is essential for precision cosmology.
Recently, we have found there notably exist observational effects on
released WMAP maps. The WMAP mission measures temperature
differences between sky points using differential radiometers
consisting of plus-horn and minus-horn$^{[1]}$. When an
antenna horn points to a sky pixel, the other one will scan a ring
in the sky with an angular radius of $141\degree$ to the centre pixel.
These measured TOD are transformed into the full-sky
temperature anisotropy map by a map-making
process$^{[2]}$. In released five-year WMAP (WMAP5) CMB maps
we find significant distortion from hot Galactic sources: the pixels
in the scan ring of a hot pixel are systematically cooled, and the
strongest anti-correlations between temperatures of a hot pixel and
its scan-ring appear at a separation angle
$\theta\sim141\degree$$^{[3]}$. The above results are confirmed
by Aurich, Lustig and Steiner$^{[4]}$. Furthermore, we also detect the
no-negligible effect of imbalance observations in published WMAP5
maps:
 systematic dependence of temperature vs. observation number difference between
the two horns of a radiometer produced by the input transmission imbalance,
and significant correlation between pixel temperature
and observation number in WMAP data$^{[5]}$. These recent findings of systematic error
in published WMAP temperature maps push us forward to further check
the WMAP map-making processing.

In this work, we find a remarkable inconsistency between the WMAP
TOD and published temperature map.
The revealed inconsistency demonstrates that there certainly exists
a serious problem in WMAP map-making process, and it is worth
checking the reliability of released WMAP results by reproducing CMB temperature
maps from the original raw data independently from the WMAP team.
We built a self-consistent software package of map-making from TOD.
Our software successfully passes a variety of tests. 
With our software, new CMB maps from WMAP TOD
are produced, which, in contrary to those published by the WMAP team, 
are well consistent with the input calibrated TOD.

\section{METHOD}
We use a simple method, the residual TOD test, to check the consistency
of WMAP map-making. Let $t_i$ denotes the temperature anisotropy at a sky pixel $i$.
In a certain band, the observed difference of the $k$th observation
$ d_k = t_{k^+} - t_{k^-}$,
where $k^+$ and $k^-$ are the sky pixels pointed by the plus-horn and minus-horn
during the observation $k$ respectively.
From  total $L$ observations,  the differential TOD
\boldmath
\begin{equation}
\label{dt1}
d = \mathbf{A}t
\end{equation}
\unboldmath
with $\mathbf{A=}\{a(k,i)\}$ being the scan matrix.
Most of elements $a(k,i)=0$ except for $a(k,i=k^+)=1$
and $a(k,i=k^-)=-1$.
The WMAP team produces the released temperature map \boldmath $\hat{t}$ \unboldmath
from the calibrated TOD  \boldmath $d$ \unboldmath by using
their map-making software$^{[2,6]}$.

For testing the consistency between the WMAP TOD
 \boldmath $d$ \unboldmath  and
 the released map  \boldmath $\hat{t}$ \unboldmath reconstructed  from
 \boldmath $d$ \unboldmath by map-making,
we calculate  for each observation $k$ the residual $d'_k$
between the measured calibrated difference and the calibrated difference predicted
by the reconstructed map, $d'_k=d_k-(\hat{t}_{k^+}-\hat{t}_{k^-})$, to get
the residual TOD
\boldmath
\begin{equation}
\label{res}
d'=d-(\hat{t}_+-\hat{t}_-)\,.
\end{equation}
\unboldmath
If the temperature map is properly reconstructed,
only the instrument noise should be left in
\boldmath $d'$ \unboldmath. To check it,
we produce the correlation map of the residual TOD
\boldmath
\begin{equation}
\label{t0} t_0=\mathbf{M}^{-1}\mathbf{A^T} d'\,,
\end{equation} \unboldmath
where  $\mathbf{M=A^TA}$  is diagonally dominant$^{[2]}$
\begin{equation}
\label{M1}
M^{-1}(i,j)\simeq \frac{1}{N_i}\delta_{ij}
\end{equation}
with $N_i$ being the total number of observations for pixel $i$.
Combining eq.~(3) and eq.~(4),  we can compute the temperature $t_0(i)$
of the correlation map of the residual TOD for each sky pixel $i$ simply by
\begin{equation}
\label{t0i}
t_0(i)=\frac{1}{N_i}(\sum_{k^+=i}d_k'-\sum_{k^-=i}d_k')\,.
\end{equation}

If the temperature map published by the WMAP team is reconstructed correctly,
the correlation  map  \boldmath $t_0$ \unboldmath
should remain only the map-making error
 with low amplitude and no significant structured signal on it.
Because eq.~(3) is linear, we have \boldmath
$t_0=\mathbf{M}^{-1}\mathbf{A^T} d - \mathbf{M}^{-1}\mathbf{A^T}
(\hat{t}_+-\hat{t}_-).$ \unboldmath For a correctly reconstructed
map  \boldmath $\hat{t}$ \unboldmath, both \boldmath ${\mathbf
M}^{-1}\mathbf{A^T} d$ \unboldmath and \boldmath $\mathbf
{M}^{-1}\mathbf{A^T} (\hat{t}_+-\hat{t}_-)$ \unboldmath will be
equal to \boldmath $\hat{t}$ \unboldmath, and then \boldmath $t_0$
\unboldmath will be exactly zero,
 despite the inevitable numerical computation error, or the map making error.

\section{RESULT}
Now we check the consistency of released temperature map with
the used WMAP TOD. 
We download  the calibrated TOD from the WMAP team's website
(fttp://lambda.gsfc.nasa.gov/).
In map-making, not all the measured TOD are used by the WMAP team.
During the preprocess of map-making, bit-coded
quality flags are set$^{[7]}$. A non-zero flag
indicates that either the observation is problematic in a specific
respect, or the beam boresight is away from one of the out planets
no more than $7\degree$, which is the antennae main beam radius
limit$^{[8]}$. In both cases, the corresponding data are at
least less optimal, and some of these data are not used
by the WMAP team in the map-making.
We follow the WMAP team's flagging convention\footnote{(1) GENERAL FLAG --
Test on bits 0, 1, 3, 4, 5. This discards data when the observatory
is not in observing mode, the Sun is visible over the shield, but includes data
when either the Earth or Moon is visible over the shield (but still in the far
sidelobes of the radiometer beams).
(2) DA-SPECIFIC FLAGS -- Test on bit 0, re-test on bits 1--10.
Bit 0 allows exclusion of data with known thermal disturbances
or radiometer upsets.  Bits 1--10 show that a planet is
close to a radiometer beam.  If bit 1--10 is set,
compute the distance between the indicated planet and the
radiometer beam centre based on the instantaneous pointings
for all points within the frame, and discard only those points
for which the planet lies within $1\degree.5$  of the beam.}
to get the used WMAP5 year-1 TOD.
Figure~1 shows the
map \boldmath $t_0$ \unboldmath
obtained by eq.~(5) from the used WMAP5 year-1 TOD and
the released WMAP5 year-1 temperatures of Q1-band, where visible structures along the
ecliptic plane and around the ecliptic poles left
 and the rms amplitude $\sim 265\,\mu$K,
which is much higher than the expected rms error of $\sim 0.2\,\mu$K
estimated by the WMAP team with flight-like simulations
for their map-making algorithm$^{[2]}$, demonstrating that 
there certainly exists a remarkable problem in the WMAP map-making process.

 \begin{figure}[t]
\begin{center}
\includegraphics[height=7cm, angle=90]{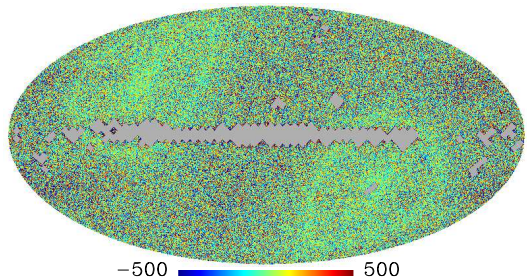}
\end{center}
\vspace{-7mm}
\caption{  Correlation maps of residual TOD  \boldmath $d'$
\unboldmath between the input TOD and the TOD predicted by the
official WMAP5 year-1 temperature map of Q1-band, in Galactic
coordinates and in units of $\mu$K. The gray area represents
pixels without valid value.} \label{fig:residual}
\end{figure}

\vspace{9mm} \begin{figure}[t]
\begin{center}
\includegraphics[height=7cm, angle=90]{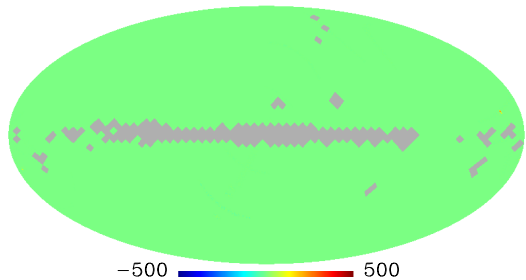}
\end{center}
\vspace{-7mm}
\caption{  Correlation map of residual TOD between the input TOD and
the TOD predicted by
 our new temperature map of Q1-band, in Galactic coordinates and in units of $\mu \rm{K}$.
The gray area represents pixels without valid value. }
\label{fig:residual2}
\end{figure}

 We then write out our map-making programs to reconstruct new 
temperature map from the used WMAP5 year-1 TOD.
According to the WMAP document$^{[2]}$, only three extra
corrections are needed:
 the $1/f$ noise removal, the sidelobe correction,
and the transmission imbalance correction. The sidelobe correction
is applied only to the K-band, and other two corrections are applied
to all bands. We perform various tests, i.e. the residual dipole
component test, the map-making convergence test, and the
end-to-end test etc, to make it sure that our data pipeline 
is in a self-consistent manner.

With our map-making program and the used WMAP5 year-1 Q1-band TOD, 
we produce new temperature map,
and with the new map and eq.~(5) we calculate the residual TOD.
The correlation map of the residual TOD is calculated and shown 
in Figure~2, where no significant
structures can be seen and the rms amplitudes is less than $0.15\,\mu$K,
almost 2000 times lower than what from the WMAP team's map-making products
and close to the rms error of  $\sim 0.11\,\mu$K for our map-making,
demonstrating that the inconsistent problem presented in \S2 can be prevented
by improving map-making algorithm. 

\section{DISCUSSION}
The remarkable structured noise left in the residual TOD of the WMAP official map 
shown in Figure 1 indicates that there must exist a serious problem 
in the map-making routine of the WMAP team, and that many previous 
cosmology studies based on the released WMAP temperature maps should be seriously affected. 
It will be very helpful if the WMAP team can thoroughly recheck their map-making 
process to find out where and how the error occurs. 
But until now they have insisted that their result has no problem, 
and prevented to discuss it with us. In view of the importance of this issue 
for cosmology study, we invite readers to explore it by themselves. 
In testing the consistency between the released map and WMAP TOD, 
the calculations for the residual TOD in eq.~(2) and its correlation map in eq.~(5) 
are considerably easy without the use of the WMAP map-making software. 
The WMAP calibrated TOD and official maps can be found 
at the website fttp://lambda.gsfc.nasa.gov/. 
 Readers can produce the used TOD from the calibrated TOD by using the WMAP team's 
flagging convention, or download the WMAP5 year-1 TOD used in this work 
from the website of Tsinghua Center for Astrophysics 
at http://dpc.aire.org.cn/data/wmap/09072731/release\_v1/tod\_for\_test/. 
	
\vspace{7mm}\begin{center}
{\bf REFERENCES}
\end{center}

\noindent 1. Bennett, C.L. et~al., 2003, ApJ, 583, 1

\noindent 2. Hinshaw, G. et~al., 2003, ApJS, 148, 63

\noindent 3. Liu, H. \& Li, T.P., 2009, Sci China G-Phy Mech Astron, 52, 804; arXiv:0809.4160v2

\noindent 4. Aurich, R., Lustig, S. \& Steiner, F., 2009, arXiv:0903.3133

\noindent 5. Li, T.P., Liu H., Song L.M., Xiong S.L. \& Nie J.Y., 2009, MNRAS, 398, 47

\noindent 6. Jarosik, N. et~al., 2007, ApJS, 170, 263

\noindent 7. Limon, M. et~al., 2008,
 {\sl Wilkinson Microwave Anisotropy Probe (WMAP) : The Five-Year Explanatory Supplement},
Greenbelt, MD: NASA/GSFC;
 Available in electronic form at http://lambda.gsfc.nasa.gov

\noindent 8. Hill, R.~S., et al. 2009, ApJS, 180, 246

\end{document}